\documentclass[onecolumn,10.5pt,compsoc]{astro-new}

\usepackage[bf]{caption}
\usepackage{cuted}  
\usepackage{epstopdf}
\usepackage{enumitem}
\usepackage[colorlinks,linkcolor=red]{hyperref}
\graphicspath{{Figures/}{figures/tem/}}
\DeclareCaptionLabelSeparator{threespace}{~\,~}
\begin{document}
\pretitle{}
\title{Dust in the Jupiter system outside the rings}
\author{Xiaodong Liu and J\"urgen Schmidt}
\addr{Astronomy Research Unit, University of Oulu, Oulu 90014, Finland}
\coremail{Xiaodong Liu, xiaodong.liu@oulu.fi}
\abstracts{Jupiter is one of the major targets for planetary exploration, and dust in the Jovian system is of great interest to researchers in the field of planetary science. In this paper, we review the five dust populations outside the ring system: grains in the region of the Galilean moons, potential dust from plumes on Europa, Jovian stream particles, particles in the outer region of the Jovian system ejected from the irregular satellites, and dust in the region of the Trojan asteroids. The physical environment for the dust dynamics is described, including the gravity, the magnetic field and the plasma environment. For each population, the dust sources are described, and the relevant perturbation forces are discussed. Observations and results from modeling are reviewed, and the distributions of the individual dust populations are shown. The understanding of the Jovian dust environment allows to assess the dust hazard to spacecraft, and to characterize the material exchange between the Jovian moons, their surface properties and distribution of non-icy constituents.}

\nomenclature
\begin{table}[!h]
\vspace {-3mm}
\begin{center}
\renewcommand{\arraystretch}{1.4}
\begin{tabular}{|p{80mm}|p{80mm}|}
\hline
$R_\mathrm{J}$&
reference radius of Jupiter (71492 km) \\
\hline
AU&
astronomical unit ($\approx 1.5\times10^8$ km)\\
\hline
DDS&
Dust Detection System\\
\hline
\end{tabular}
\label{tab1}
\end{center}
\vspace {-6mm}
\end{table}

\section{Introduction}
\noindent \textbf{Interplanetary and circumplanetary} dust particles have typical sizes ranging from nanometers up to millimeters. Dust is widely distributed in the Jovian system, \textbf{where particles were first detected by the Pioneer spacecraft in the early 1970s \cite{humes1974interplanetary,humes1975pioneer}, and the dusty ring system was first observed by the cameras onboard the Voyager spacecraft in 1979 \cite{smith1979jupiter, owen1979jupiter}. Subsequently, dust particles in the Jovian system were detected in-situ by space missions, and dust rings were observed by ground- and space-based telescopes as well as cameras onboard spacecraft.} So far, eight spacecraft have made measurements/observations of dust particles/rings in the Jupiter system, including Pioneer 10 and 11 \cite{humes1974interplanetary,humes1975pioneer,zeehandelaar2007local}, Voyager 1 and 2 \cite{smith1979jupiter, owen1979jupiter, 1985Natur.316..526S,1987Icar...69..458S, throop2004jovian, showalter2008properties}, Galileo \cite{ockert1999structure, burns1999formation, 2000Icar..146....1M, 2004Icar..170...35B, throop2004jovian, showalter2008properties, 1998P&SS...47...85K, 2001P&SS...49.1285K,kruger2003impact, Kruger:2006jeb, 2009Icar..203..198K, 2010P&SS...58..965K, grun1996constraints, Thiessenhusen:2000iq,krivov2002tenuous,krivov2002dust}, Cassini \cite{throop2004jovian, 2003Sci...299.1541P,2003Icar..164..461B}, New Horizons \cite{Showalter:2007km, 2010GeoRL..3711101P}, and Ulysses \cite{grun1992ulysses,1993Natur.362..428G,2006P&SS...54..919K}. Besides, the Hubble Space Telescope \cite{1999Icar..141..253M,showalter2008properties} and the ground-based Keck telescope \cite{de1999keck,showalter2008properties, de2008keck} were used to observe the Jovian ring system.

\begin{figure}
   \centering
   \includegraphics[width=16cm,angle=0]{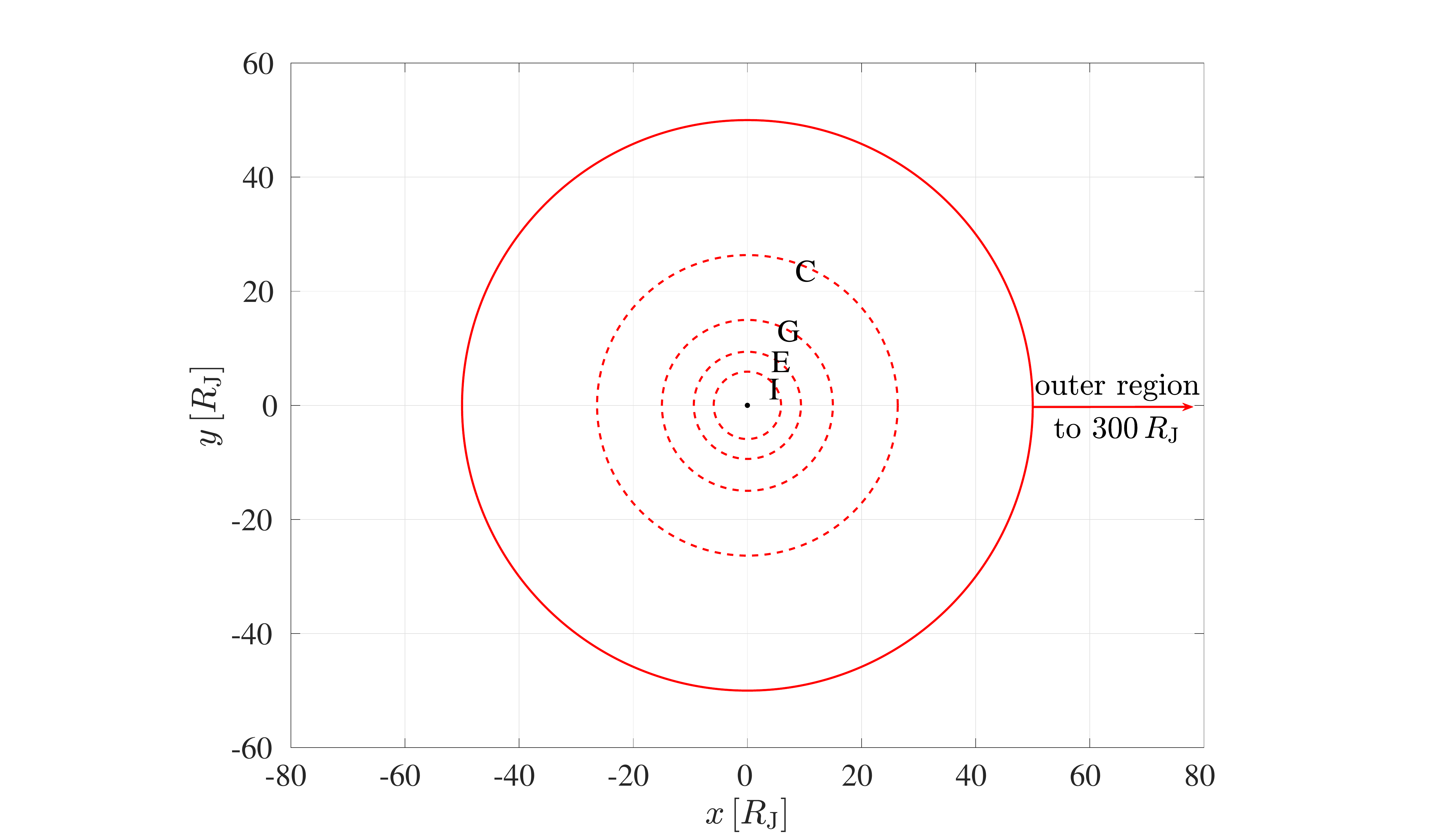}
   \caption{Top view showing the orbits of the four Galilean moons and the outer region. Jupiter lies \textbf{at} the origin. The orbits of Io (5.9 $R_\mathrm{J}$), Europa (9.4 $R_\mathrm{J}$), Ganymede (15.0 $R_\mathrm{J}$) and Callisto (26.3 $R_\mathrm{J}$) are denoted as I, E, G, and C, respectively. The outer region of the Jovian system extends over a range of [50, 300] $R_\mathrm{J}$.}
          \label{fig:galilean_outer_region_sketch}
   \end{figure}

\begin{figure}
   \centering
   \includegraphics[width=16cm,angle=0]{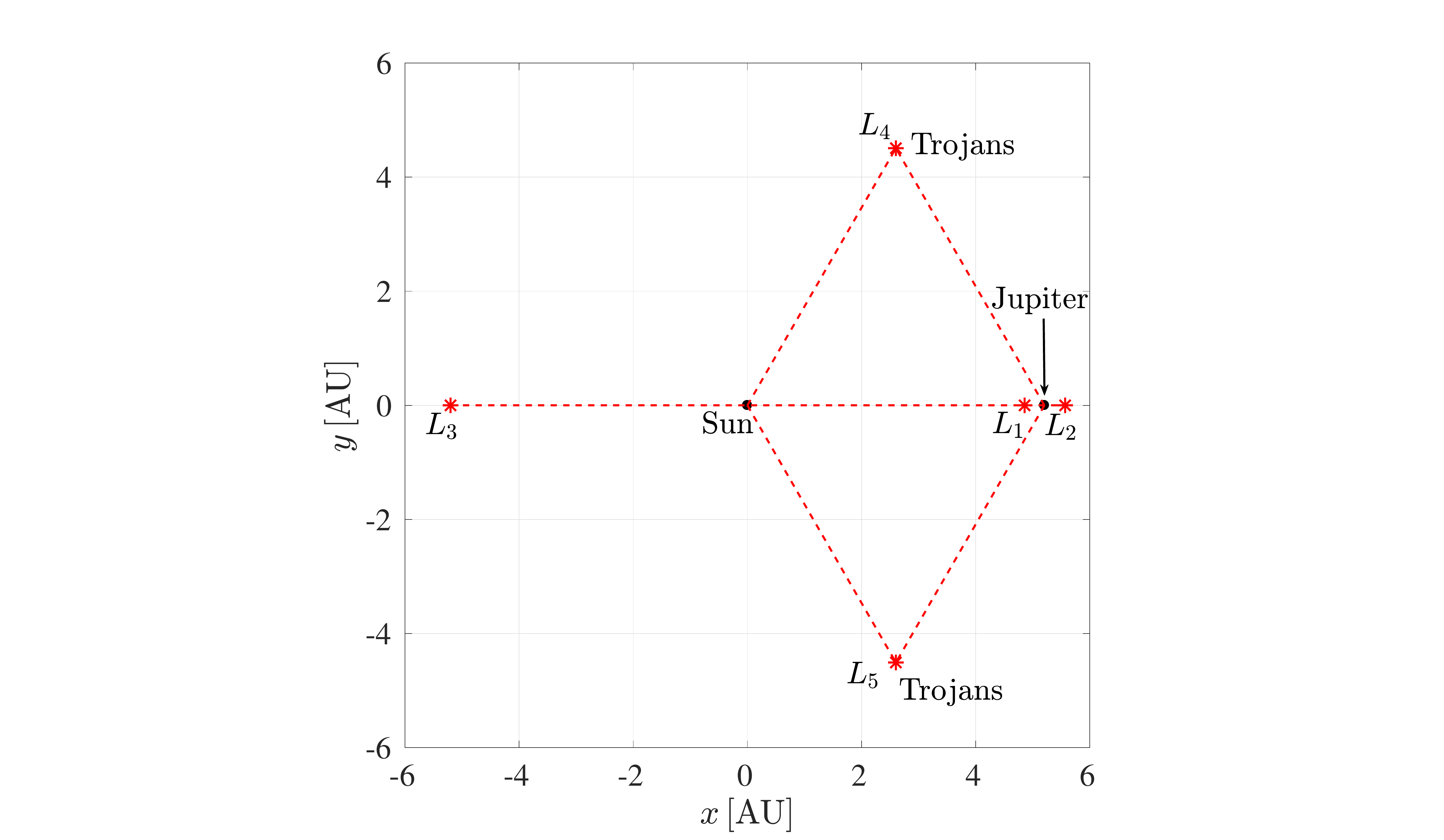}
   \caption{Locations of the Jovian Trojan asteroids. The Sun lies at the origin. Jupiter is about 5.2 AU from the Sun. The Lagrange points in the Sun-Jupiter system are indicated as red asterisks, including three collinear Lagrange points ($L_1$, $L_2$, $L_3$) and two triangular Lagrange points ($L_4$, $L_5$). The $L_4$ and $L_5$ Trojan asteroids librate around the triangular Lagrange points $L_4$ and $L_5$, respectively.}
          \label{fig:trojan_sketch}
   \end{figure}

Dust particles in the Jovian system \textbf{(including the Sun-Jupiter system)} are roughly divided into the following six different populations: (1) particles in the Jovian ring system including the main ring, the halo ring and two gossamer rings \cite{smith1979jupiter, owen1979jupiter, burns1999formation}, (2) \textbf{grains in the region of the Galilean moons \cite{krivov2002tenuous,Thiessenhusen:2000iq}, which derive from impact-generated dust clouds around the Galilean moons \cite{kruger1999detection,kruger2003impact} as well as magnetospherically captured interplanetary and interstellar particles \cite{colwell1998capture, colwell1998jupiter,Thiessenhusen:2000iq}}, (3) possible dust from plumes on Europa \cite{quick2013constraints, roth2014transient, roth2014orbital, southworth2015modeling, sparks2016probing}, (4) small and fast dust streams (nanometer-sized, $>200 \, \mathrm{km\,s^{-1}}$) from Io's volcanic plumes \cite{1993Natur.362..428G,grun1996constraints,2000Natur.405...48G}, (5) particles in the outer region from the Jovian irregular moons \cite{krivov2002dust}, \textbf{and (6) dust grains in the region of Jupiter's Trojan asteroids \cite{liu2018dust, zimmer2014orbital, de2010studying}. Previous studies on the first dust population, i.e.~the Jovian ring system are well covered by several reviews \cite{de2018rings,burns2004jupiter,miner2007planetary}, including a recent one published in 2018 \cite{de2018rings}, to which the reader is referred. Thus, in the current review paper we put more emphasis on the latter five populations} (see Figure \ref{fig:galilean_outer_region_sketch} for the locations of the four Galilean moons and the outer region of the Jovian system, and Figure \ref{fig:trojan_sketch} for the locations of the Trojan asteroids associated with the Lagrange points in the Sun-Jupiter system). \textbf{The reader is also referred to an earlier review from 2004 \cite{kruger2004jovian}, by which the dust populations (2), (4) and (5) were well covered.}
   
In the near future, there are two reconnaissance missions that will visit the Jovian system. One is the JUpiter ICy moons Explorer mission \cite{plaut2014jupiter} developed by European Space Agency (ESA), which is scheduled for launch in 2022. The Europa Clipper mission \cite{phillips2014europa}, developed by National Aeronautics and Space Administration (NASA), is planned to launch between 2022-2025. The SUrface Dust Analyzer, a dedicated dust mass spectrometer instrument, is onboard Europa Clipper. This dust detector will measure the composition of the dust particles lofted from the Galilean moons. The Lucy mission developed also by NASA will visit Jupiter's $L_4$ and $L_5$ Trojan asteroids \cite{levison2017lucy}, which is scheduled for launch in 2021.

\section{Physical environment for dust dynamics}
\noindent The dynamics of the dust particles in circumjovian space is influenced by various forces, including Jupiter's gravity, the Lorentz force, solar radiation pressure, Poynting-Robertson drag, plasma drag, and the gravity from other celestial bodies. Taking into \textbf{account} all these forces, the equation of motion of a dust grain in the Jovicentric inertial frame \textbf{reads} \cite{burns1979radiation, kruger2004jovian, liu2016dynamics}
\begin{equation} \label{equ_dynamic_model_circumjovian}
\begin{split}
\ddot{\vec r} = & GM_\mathrm J\nabla\left\{\frac{1}{r}\left[1-\sum_{n=1}^{N_J}J_{2n}\left(\frac{R_\mathrm J}{r}\right)^{2n}P_{2n}(\cos\theta)\right]\right\}   +   \frac{Q}{m_\mathrm{g}}\left(\dot{\vec r}-\vec \Omega_\mathrm{J}\times \vec r\right)\times{\vec B}   \\
& + \frac{3Q_\mathrm SQ_\mathrm {pr}\mathrm{AU}^2}{4(\vec r-{\vec r}_\mathrm S)^2\rho_\mathrm gr_\mathrm gc}\left\{\left[1-\frac{(\dot{\vec r}-\dot{\vec r}_\mathrm S)\cdot\hat{\vec r}_\mathrm {Sd}}{c}\right]\hat{\vec r}_\mathrm {Sd}   -   \frac{\dot{\vec r}-\dot{\vec r}_\mathrm S}{c}  \right\}   \\
& +\vec F_\mathrm{direct\_drag}   +   \vec F_\mathrm{Coulomb\_drag}   +   Gm_\odot\left(\frac{\vec r_\mathrm{dS}}{r_\mathrm{dS}^3}-\frac{\vec r_\mathrm S}{r_\mathrm S^3}\right)   +   \sum_{i=1}^{N_\mathrm{m}}Gm_{\mathrm m_i}\left(\frac{\vec r_{\mathrm {dm}_i}}{r_{\mathrm {dm}_i}^3}-\frac{\vec r_{\mathrm m_i}}{r_{\mathrm m_i}^3}\right) \,.
\end{split}
\end{equation}
Here \textbf{$\vec r$ is the Jovicentric position vector of the particle, $G$ is the gravitational constant}, $M_\mathrm J$ is the mass of Jupiter, $J_{2n}$ is the even degree zonal of spherical harmonics for Jupiter's gravity field, $N_J$ is the number of $J_{2n}$, $P_{2n}$ is the Legendre polynomial of degree $2n$, $\theta$ is the colatitude in the body-fixed frame, \textbf{$Q$ is the grain charge, $m_\mathrm{g}$ is the mass of the grain, $\vec B$ is the local Jovian magnetic field}, $\vec \Omega_\mathrm{J}$ is the angular velocity of Jupiter's rotation, \textbf{$Q_\mathrm S$ is the solar radiation flux density at a heliocentric distance of one AU, $Q_\mathrm {pr}$ is the radiation efficiency factor, $\rho_\mathrm g$ is the bulk density of the grain, $r_\mathrm g$ is the particle radius, $c$ is the light speed}, ${\vec r}_\mathrm S$ is the position vector of the Sun, $\hat{\vec r}_\mathrm {Sd}$ is the unit vector along the radiation flux, \textbf{$m_\odot$ is the mass of the Sun}, $\vec r_\mathrm{dS}$ is the vector from the particle to the Sun, $N_\mathrm{m}$ is the number of the moons that are included as gravitational perturbers, $m_{\mathrm m_i}$ is the mass of the $i$th moon, $\vec r_{\mathrm {dm}_i}$ is the vector from the particle to the $i$th moon, and $\vec r_{\mathrm m_i}$ is the position vector of the $i$th moon. 
The expression for the direct plasma drag $\vec F_\mathrm{direct\_drag}$ is described in \cite{1979ApJ...231...77D, banaszkiewicz1994evolution}, and its supersonic approximation and subsonic approximation can be found in \cite{morfill1980dust, dikarev1999dynamics}. The expression for the Coulomb drag force $\vec F_\mathrm{Coulomb\_drag}$ is given in \cite{1979ApJ...231...77D, northrop1990plasma}.

First, we discuss the gravitational environment for dust particles in circumjovian space. Because Jupiter rotates rapidly, it has a shape of an oblate spheroid. As a result, the even zonals of spherical harmonics in the gravity field are the dominant perturbations of the central gravity field. The gravitational perturbations from the four inner moons (Adrastea, Metis, Amalthea and Thebe), four Galilean moons (Io, Europa, Ganymede and Callisto), and the irregular moons are important for dust evolving in their orbital ranges. \textbf{The} solar gravitational perturbation is important for dust grains from the irregular moons.

The Jovian magnetic environment was studied by many researchers, which is needed to calculate the Lorentz force. Important recent models include the O6 model \cite{connerney1993magnetic}, the Ulysses model \cite{1996JGR...10124929D}, the VIP4 model \cite{1998JGR...10311929C}, the Amalthea model \cite{1998JGR...10317535R}, the VIT4 model \cite{connerney2007planetary}, and the VIPAL model \cite{hess2011model}. The Jovian magnetic field is not a pure dipole field. Especially inside the Jovian ring system, the quadrupole and octupole terms are important for the dust dynamics \cite{schaffer1987dynamics,burns1985lorentz}.

The ambient plasma affects the dust dynamics in terms of dust charging and plasma drag. Many plasma models have been developed \cite{divine1983charged, sittler1987io, frank2002galileo, bagenal2011flow, garrett2015jovian}. Here we discuss the frequently cited global plasma models, the DG1 model \cite{divine1983charged} and its updated version the DG2 model \cite{garrett2015jovian}. The DG1 model \cite{divine1983charged} was mainly based on the data from the Pioneer and Voyager spacecraft. The DG2 model \cite{garrett2015jovian} was updated by data from the Voyager spacecraft and the Galileo spacecraft \cite{sittler1987io, frank2002galileo, bagenal2011flow}. According to these two models, the Jovian plasma environment is divided into three populations: radiation belt electrons and protons (0.1-100 MeV), warm plasma (0.1-100 keV), and cold electrons, protons and heavy ions (1-100 eV). The cold heavy ion species include $\mathrm{O}^+$, $\mathrm{O}^{++}$, $\mathrm{S}^+$, $\mathrm{S}^{++}$, $\mathrm{S}^{+++}$ and $\mathrm{Na}^+$ \cite{divine1983charged,garrett2015jovian}, which are the most important species for plasma drag. In the Jovian plasma environment, the dust particles are charged \cite{horanyi1996charged}. The charging currents include electron and ion currents, secondary electron emission and photoelectron emission \cite{horanyi1996charged}. For the small stream particles, the dust surface potential is size-dependent \cite{dzhanoev2016charging} because of the small-size effect of the secondary electron emissions. For small grains the charging becomes a stochastic process, dealing with absorption of single unit charges.

For dust particles in the region of the Trojan asteroids, the dominant forces are solar gravity, solar radiation pressure, solar wind drag, Poynting-Robertson drag, the solar wind induced Lorentz force, and gravitational perturbations from Jupiter and other planets. The equation of motion of Trojan dust particles in the heliocentric inertial frame can be expressed as \cite{burns1979radiation, gustafson1994physics, liu2018dust}
\begin{equation} \label{equ_dynamic_model_trojan}
\begin{split}
\ddot{\vec r}_\odot = &-\frac{Gm_\odot}{r_\odot^3}{\vec r_\odot} + \sum_{i=1}^{N_\mathrm{P}}GM_{\mathrm P_i}\left(\frac{\vec r_{\mathrm {dP}_i}}{r_{\mathrm {dP}_i}^3}-\frac{\vec r_{\mathrm P_i}}{r_{\mathrm P_i}^3}\right) + \frac{Q}{m_\mathrm{g}}\left(\dot{\vec r}_\odot-\vec v_{sw} \right)\times{\vec B} \\
& + \frac{3Q_\mathrm SQ_\mathrm {pr}\mathrm{AU}^2}{4r_\odot^2\rho_\mathrm gr_\mathrm gc}\left\{\left[1-(1+sw)\frac{\dot r_\odot}{c}\right]\hat{\vec r}_\odot - (1+sw)\frac{\dot{\vec r}_\odot}{c}\right\} \,.
\end{split}
\end{equation}
Here, \textbf{$\vec r_\odot$ is the heliocentric position vector of the particle}, $N_\mathrm{P}$ is the number of the planets that are included as gravitational perturbers, $M_{\mathrm P_i}$ is the mass of the $i$th planet, $\vec r_{\mathrm {dP}_i}$ is the vector pointing from the grain to the $i$th planet, $\vec r_{\mathrm P_i}$ is the position vector of the $i$th planet, \textbf{$\vec B_\odot$ is the local interplanetary magnetic field}, $\vec v_{sw}$ is the velocity of the solar wind, and $sw$ defines the ratio of solar wind drag relative to the Poynting-Robertson drag. 

The Sun, Jupiter and the Trojan particle form the traditional three-body problem. The solar gravity coupled with Jupiter's gravity tends to make the particles move as their sources in tadpole orbits. The Lorentz force due to the solar wind is induced by the charged particles moving in the interplanetary magnetic field. A description of a model for the interplanetary magnetic field can be found in \cite{gustafson1994physics, landgraf2000modeling}.

\section{Dust in the region of the Galilean moons}
\begin{figure}
   \centering
   \includegraphics[width=10cm,angle=0]{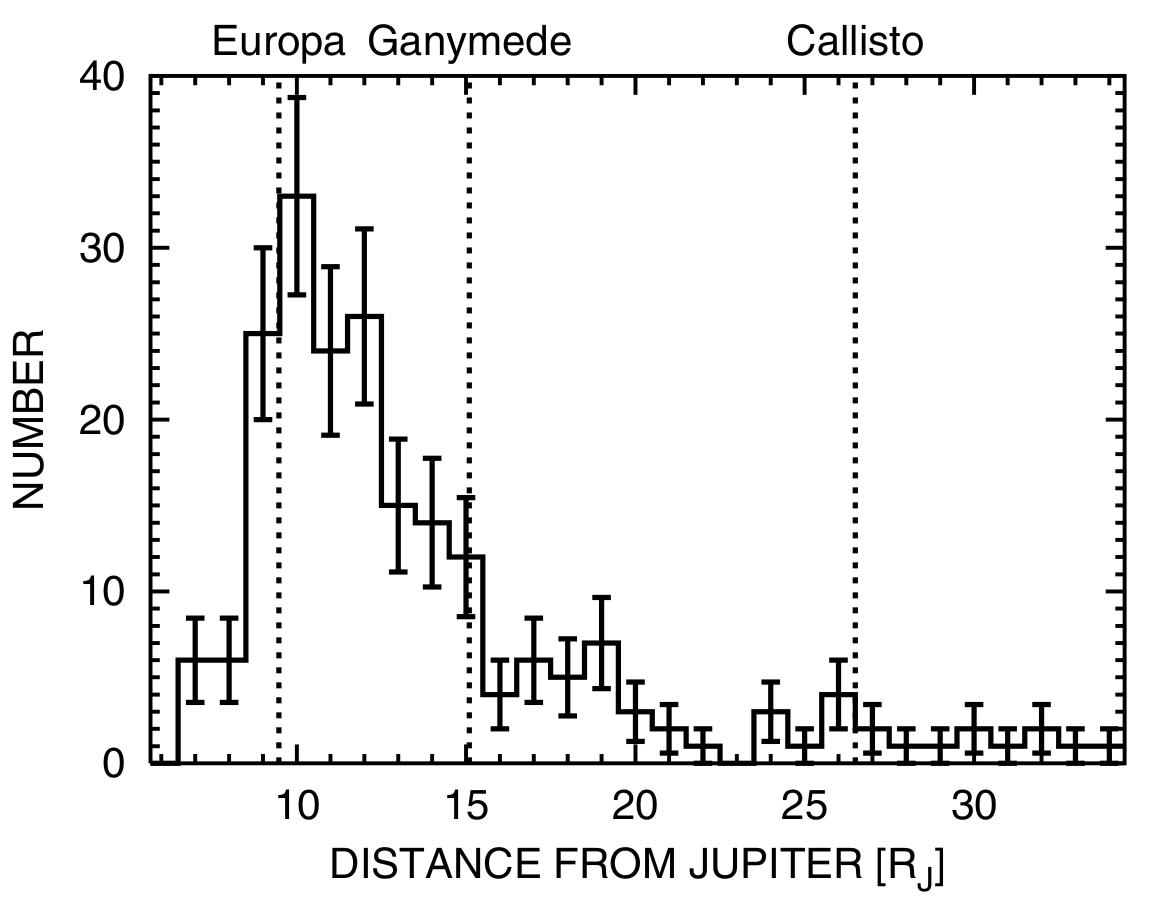}
   \caption{\textbf{Number of dust particles detected by the Galileo DDS in the second amplitude range AR2 (from the year 1996 to 2001) vs.~the Jovicentric distance. The error bars were overplotted. The dashed lines denote the location of the Galilean moons Europa, Ganymede and Callisto. Reprinted from \cite{krivov2002tenuous}, \copyright 2002, American Geophysical Union, with permission from John Wiley and Sons.}}
          \label{fig:Galilean_num_AR2}
   \end{figure}

\begin{figure}
   \centering
   \includegraphics[width=10cm,angle=0]{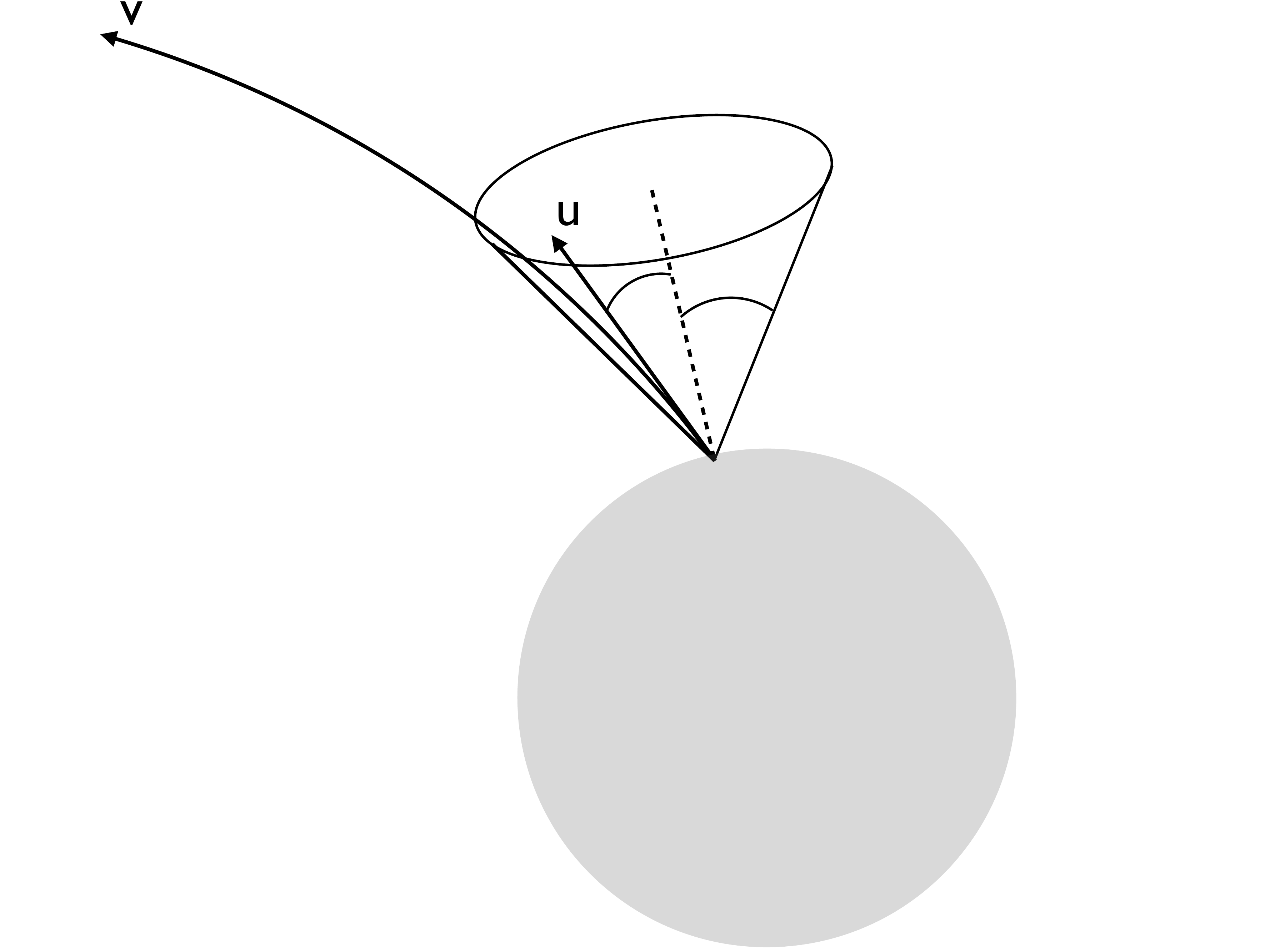}
   \caption{Schematic diagram of the impactor-ejecta process. A fast interplanetary impactor hits the surface of an atmosphereless body and releases ejecta within a cone from the surface. \textbf{The symbol $\mathbf{u}$ denotes the initial velocity of a particle at the ejection point on the surface of the source body, and $\mathbf{v}$ denotes the velocity of this particle at distance. The arrows indicate the directions of $\mathbf{u}$ and $\mathbf{v}$, respectively.} A fraction of the ejecta can escape the gravity of the body and evolve into circumplanetary/circumsolar space, and other ejecta fall back to the body's surface. This mechanism applies to atmosphereless bodies including planetary moons \cite{kruger1999detection,horanyi2015permanent} and asteroids \cite{szalay2016impact}.}
          \label{fig:JSsketch2}
   \end{figure}

\noindent Dust particles in the region of the Galilean moons were detected by the DDS onboard the Galileo spacecraft \cite{KRUGER2002144, krivov2002tenuous,Thiessenhusen:2000iq} \textbf{(see Figure \ref{fig:Galilean_num_AR2})}, for which the Galilean moons Io, Europa, Ganymede and Callisto are likely the dominant sources \cite{krivov2002tenuous,Thiessenhusen:2000iq}. Dust particles are ejected from the surfaces of these moons by hypervelocity impacts of interplanetary micrometeoroids (see Figure \ref{fig:JSsketch2} for the illustration of this process). These ejecta form dust clouds in the vicinity of the Galilean moons \cite{kruger1999detection, 2000P&SS...48.1457K, kruger2003impact}.

The initial size distribution of the dust ejecta follows a power law
\begin{equation}
p(r_\mathrm{g})\propto r_\mathrm{g}^{-q} \,.
\end{equation}
The exponent $q$ is the slope of the differential production rate of the dust ejecta. The value $q=3.4$ is consistent with measurements of the dust clouds around the Galilean moons by the Galileo DDS \cite{2000P&SS...48.1457K, kruger2003impact}. A steeper value $q=3.73$ is indicated by measurements of the dust cloud enshrouding the Earth Moon by the Lunar Dust Experiment (LDEX) instrument onboard the Lunar Atmosphere and Dust Environment Explorer (LADEE) spacecraft \cite{horanyi2015permanent}.

A fraction of the ejected particles escape into circumjovian space, and are distributed in the region of the Galilean moons. Magnetospherically captured interplanetary and interstellar particles also contribute to the dust population in this region \cite{colwell1998capture, colwell1998jupiter,Thiessenhusen:2000iq}. Besides, by analyzing the Galileo DDS data \cite{soja2015new}, it was inferred that a fraction of dust particles in this region may come from grains transported inward from Jupiter's outer irregular moons \cite{bottke2013black}, particles transported outward from Jupiter's gossamer rings, as well as focused interplanetary and interstellar dust. Moreover, cometary dust can be captured as it was for comet Shoemaker-Levy 9 \cite{horanyi1994new}.

The detected dust number density in the region of the Galilean moons is in the order of $10^2$-$10^3$ $\mathrm{km}^{-3}$ \cite{2009Icar..203..198K, 2013P&SS...75..117S, krivov2002tenuous, KRUGER2002144, Thiessenhusen:2000iq}. The fate of dust particles from the three Galilean moons Europa, Ganymede and Callisto were investigated in \cite{krivov2002tenuous}. The dust production rates were estimated, and the orbital behavior of submicron- and micron-sized particles was analyzed. It was shown that there exists a tenuous ring in the region of the Galilean moons and simulations were compared to the Galileo DDS data. Particles sized 5 $\mu\mathrm{m}$ from the surfaces of the Galilean moons were integrated, leading to a model that gives a reasonable fit to the dust detection by the Pioneer 10 and 11 spacecraft \cite{zeehandelaar2007local}. Later, a semi-analytical model was developed to analyze the dynamics and distribution of charged particles in the region of the Galilean moons \cite{sachse2017planetary}. High accuracy simulations for a large number of particles of various sizes were performed in \cite{liu2016dynamics}, including the gravitational $J_2$, $J_4$ and $J_6$ terms, Lorentz force, solar radiation pressure, Poynting-Robertson drag, plasma drag, and the gravitational perturbations from the Sun and the four Galilean moons; the particle lifetimes, the asymmetry of the dust configuration in the frame rotating with the Sun, the transport of dust between the Galilean moons and to Jupiter, as well as the distribution of orbital elements of the dust grains were investigated.
   
In the region of the Galilean moons there exists an appreciable fraction of particles on retrograde orbits detected by the Galileo DDS \cite{Thiessenhusen:2000iq, 2013P&SS...75..117S}. The sources of these particles are (1) captured interplanetary and interstellar dust through exchange of energy and angular momentum with the Jovian magnetosphere \cite{colwell1998capture, colwell1998jupiter,Thiessenhusen:2000iq}, (2) particles transported inward from the Jovian outer irregular retrograde moons \cite{soja2015new, bottke2013black}, (3) particles from comet Shoemaker-Levy 9 \cite{horanyi1994new}, and (4) grains from Galilean moons with large inclinations excited by solar radiation pressure and Lorentz force \cite{liu2016dynamics}.

\section{Possible dust plumes on Europa}
\begin{figure}
   \centering
   \includegraphics[width=10cm,angle=0]{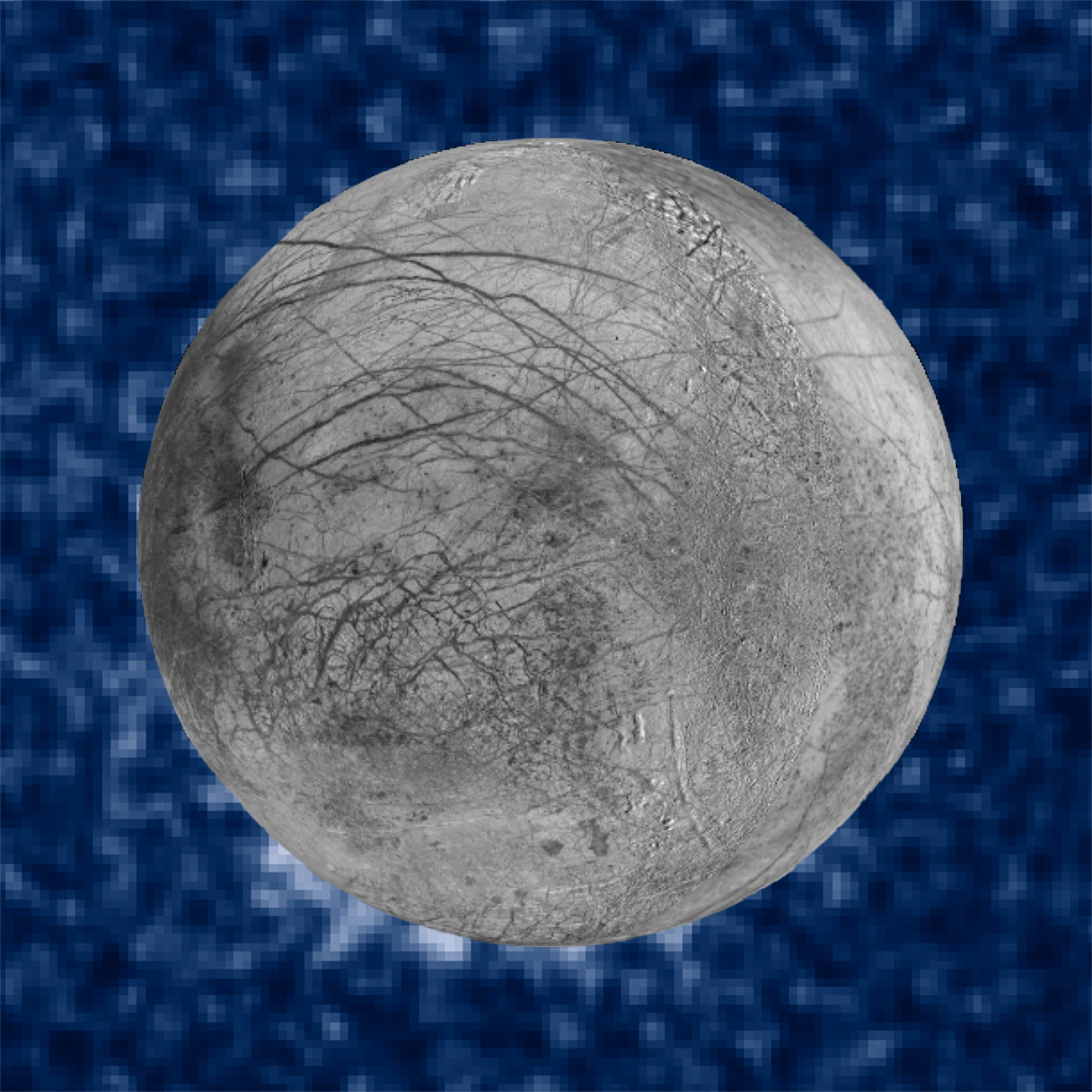}
   \caption{The suspected Europa's water vapor plume is visible in the lower left of the composite image. The image of the plume was taken by Hubble's Imaging Spectrograph on January 26, 2014. The image of Jupiter's moon Europa was obtained from data returned by Galileo and Voyager. Credit: \href{https://www.nasa.gov/}{NASA}, \href{http://www.spacetelescope.org/}{ESA}, W. Sparks (STScI), and the USGS Astrogeology Science Center.}
          \label{fig:europa_plume}
   \end{figure}

\noindent Europa is the major scientific target of the Europa Clipper mission, and will also be investigated by the JUpiter ICy moons Explorer mission. Because of its potential habitability, Europa received much interest by researchers in recent years, including the plume activity on Europa (Figure \ref{fig:europa_plume}). In 2012, a water vapor plume was detected in Europa's southern hemisphere for the first time when the satellite was close to orbital apocenter \cite{roth2014transient}. Follow-up observations showed that orbital location of Europa is not a sufficient condition to detect its plume \cite{roth2014orbital}. Further analysis \cite{rhoden2015linking} showed that the orbital location of Europa at true anomaly of about $120^{\circ}$ is a favorable position to detect the plume. Later observations showed that the plume was also detected near Europa's equatorial regions \cite{sparks2016probing}.

\begin{figure}
\centering 
\includegraphics[width=16cm]{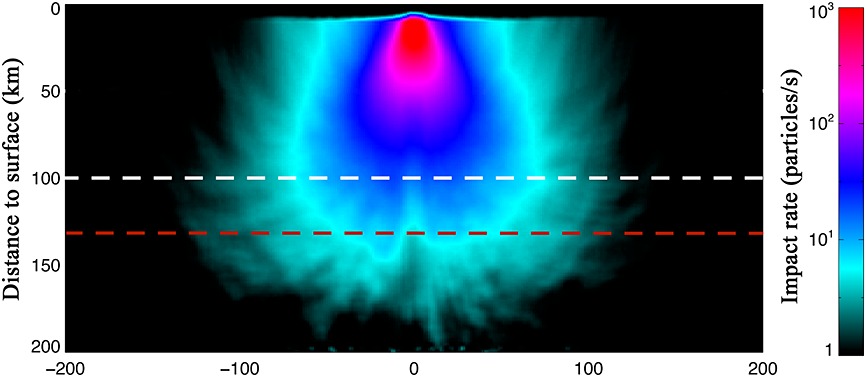}
\caption{\textbf{Vertical impact rate of Europa plume dust that can be recorded by the dust instrument onboard spacecraft. The result was obtained from simulations. The white dashed line denotes the favorable upper altitude of spacecraft, and the red dashed line denotes a parameter-specific larger bound. Reprinted from \cite{southworth2015modeling}, \copyright 2015, American Geophysical Union, with permission from John Wiley and Sons.}}
\label{fig:Plume_dust}
\end{figure}

Europa's plume may contain dust particles \cite{quick2013constraints, southworth2015modeling}, as it is the case for Saturn's moon Enceladus. The Europa's dust plume was modeled in \cite{southworth2015modeling}, including Jupiter's gravity with $J_2$ term, as well as Europa's gravity. The dust plume structure in terms of the three-dimensional particle number density was derived, \textbf{and the dust impact rate that can be recorded by the dust instrument onboard spacecraft was calculated \cite{southworth2015modeling} (Figure \ref{fig:Plume_dust}).} The Europa dust plume was also compared with the Enceladus plume, and the differences were shown. These results are useful to design the SUrface Dust Analyzer instrument onboard the Europa Clipper spacecraft. As it is the case for the Enceladus plume \cite{2009Natur.459.1098P, 2011Natur.474..620P}, the sampling of plume material allows to directly constrain the interior composition of Europa.

\section{Jovian stream particles}
\begin{figure}
   \centering
   \includegraphics[width=15cm,angle=0]{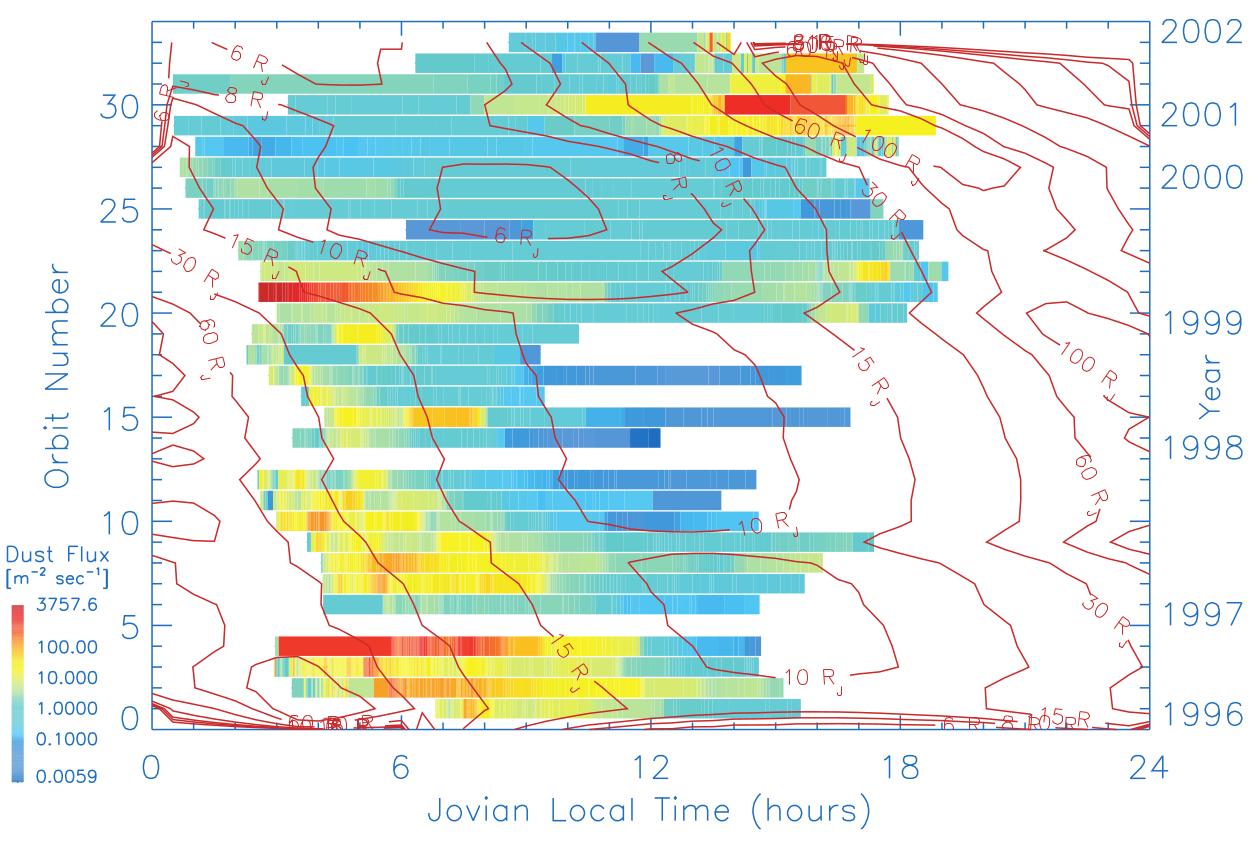}
   \caption{\textbf{Measured stream fluxes by the Galileo DDS from the year 1996 to 2002 (along 33 spacecraft orbits around Jupiter) vs.~the Jovian local time. The solid line in red denotes the Jovicentric distance $r$. The values of the stream fluxes were scaled by multiplying by a factor of $(r/6R_\mathrm{J})^2$. Reprinted from \cite{2003GeoRL..30b..30K}, \copyright 2003, American Geophysical Union, with permission from John Wiley and Sons.}}
          \label{fig:Stream_fluxes}
   \end{figure}

\noindent \textbf{Jovian} stream particles are tiny nanometer-sized dust particles that leave the Jovian system at high speed ($>200 \, \mathrm{km\,s^{-1}}$), \textbf{and their fluxes are highly variable (see Figure \ref{fig:Stream_fluxes}; also see Figure 1 in \cite{Kruger:2003bj} and Figure 8 in \cite{kruger2005galileo})}. The stream particles were first discovered by the Ulysses spacecraft when approaching Jupiter at a large distance from the planet \cite{1993Natur.362..428G, 1996Sci...274.1501Z}, and thus were believed to originate from the Jovian system. The detection of the stream particles also by the Galileo spacecraft confirmed the Jovian origin \cite{grun1996constraints}. A periodicity in the impact rate of the stream particles recorded by the Galileo DDS was found to compatible with Io's orbital period, which implies that Io is the most likely source for the stream particles \cite{1999Ap&SS.264..247K, 2000Natur.405...48G}. This was confirmed by their correlation with Io's volcanic plume activity \cite{Kruger:2003bj}, and by the variation of the stream flux with the Jovian local time \cite{2003GeoRL..30b..30K} \textbf{(Figure \ref{fig:Stream_fluxes})}, as well as by the compositional analysis of the stream particles performed by the Cosmic Dust Analyzer (CDA) onboard the Cassini spacecraft \cite{2006Icar..183..122P}. This compositional analysis \cite{2006Icar..183..122P} identified sodium chloride (NaCl) as the dominant constituent of the stream particles, with contribution from potassium and sulphur/sulphurous compounds, and possibly a small fraction of silicates or rocky minerals.

The surface equilibrium potential for the Jovian stream particles depends strongly on grain size \cite{dzhanoev2016charging}. Due to their small size and the dependence of the Lorentz force on the inverse square of the grain size, their trajectories are strongly affected by the Lorentz force induced by the Jovian magnetic field \cite{horanyi1993mechanism, horanyi1993dusty, 1997GeoRL..24.2175H, 1998JGR...10320011G} and by the interplanetary magnetic field \cite{1993Natur.362..428G, hamilton1993ejection, grun1996constraints, 1996Sci...274.1501Z}. The stability of small dust grains in the Jovian magnetic field were analyzed, and the stability boundaries for these small particles were evaluated \cite{jontof2012fate_a, jontof2012fate_b}. The charging of the stream particles in Io's cold plasma torus was discussed, where the stream particles are charged negatively by the capture of the cold electrons \cite{2004GeoRL..3116802F}. In that study, the maximum size of stream particles that can escape from Io was also calculated. The escaping stream particles get recharged positively in the Jovian magnetosphere due to the strong secondary electron current and are eventually ejected from the Jovian system by the Lorentz force \cite{horanyi1993mechanism, horanyi1993dusty}. The analysis of the Ulysses data showed that the observed pattern for the stream particles is closely correlated with the compression regions in the interplanetary magnetic field (regions of enhanced magnetic field) within a few AU from Jupiter \cite{flandes2011magnetic}. 

The stream particles may be a significant source for non-icy material on the Galilean moons. Therefore, their characterization is important to understand the contribution of other mechanisms that bring non-icy material to the surface from the interior of the moons.

\section{Dust in the outer region of the Jovian system}
\noindent In the outer region of the Jovian system (about [50, 300] $R_\mathrm{J}$ from Jupiter), there exists a dust population with a nearly constant number density of about 10 $\mathrm{km}^{-3}$, which was detected by the Galileo DDS \cite{krivov2002dust}. The primary source of this dust population is believed to be the Jovian outer irregular moons, which generate dust particles by impacts of interplanetary micrometeoroids. The dominant perturbation forces for particles in this region are solar radiation pressure, Poynting-Robertson drag, solar gravity, and the Lorentz forces induced by the Jovian magnetic field and the interplanetary magnetic field. The dynamics of dust particles in this region was analyzed for submicron- and micron-sized particles by including solar radiation pressure, Poynting-Robertson drag and solar gravity \cite{krivov2002dust}. The fate of large debris from the outer, irregular moons was studied in \cite{bottke2013black}. In their simulations, the motion of particles in the range of [5, 200] $\mu\mathrm{m}$ (diameter) influenced by Jupiter's oblateness, solar radiation pressure, Poynting-Robertson drag and planetary perturbations were integrated, and the efficiency and timescale of particle accretion on the Galilean moons Callisto, Ganymede, Europa and Io was analyzed. It was suggested that the dark color of the surfaces of the outer three Galilean moons is primarily due to accumulation of these particles from the outer irregular moons \cite{bottke2013black}. The accumulation rates could be highest for Callisto, intermediate for Ganymede, and least for Europa, explaining the gradient in coloration of the surfaces of these moons.

\section{Trojan dust}
\noindent Trojan asteroids consist of two swarms of asteroids that share the orbit of Jupiter. At the time of this writing, there are more \textbf{than} four thousand Jovian $L_4$ Trojan asteroids and more than two thousand $L_5$ Trojan asteroids that have been discovered (Jet Propulsion Laboratory Small-Body Database Search Engine (\url{https://ssd.jpl.nasa.gov/sbdb_query.cgi}); IAU Minor Planet Center (\url{https://www.minorplanetcenter.net/iau/lists/JupiterTrojans.html})). To date there was no spacecraft mission to conduct reconnaissance of the Trojan swarms. The Lucy spacecraft (scheduled for launch in 2021) will be the first mission to fly by the Jupiter Trojan asteroids \cite{levison2017lucy}.

\begin{figure}
\centering 
\includegraphics[width=11cm]{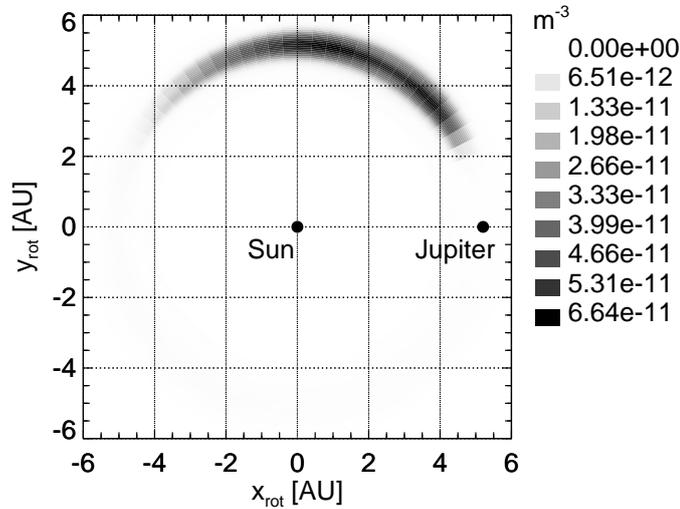}
\caption{\textbf{Number density of dust particles ejected from the $L_4$ Trojan asteroids in the frame rotating with Jupiter. The result was calculated from long-term simulations for particles larger than $0.5 \, \mathrm{\mu m}$, and was averaged vertically over $[-0.55, 0.55] \, \mathrm{AU}$. The Sun lies at the origin, and Jupiter lies at the positive $x$ axis with a heliocentric distance of about $5.2 \, \mathrm{AU}$. Credit: Liu and Schmidt, A\&A, 609, A57, 2018 \cite{liu2018dust}, reproduced with permission from Astronomy \& Astrophysics, \copyright ESO.}}
\label{fig:SimuPolar}
\end{figure}

There exist only few studies of dust particles in the region of Jupiter Trojan asteroids. The non-detection of infrared brightness by the Cosmic Background Explorer (COBE) satellite \cite{kuchner2000search} gives an upper limit of $6\times10^{13} \, \mathrm{m^2}$ for the total cross section of particles roughly larger than 10 $\mu\mathrm{m}$ (radius) in the region of the $L_5$ Trojans \cite{jewitt2000population}. The dominant dust creation mechanisms are interplanetary micrometeoroid impacts on the surfaces of the Trojan asteroids \cite{liu2018dust}. In previous research \cite{liu2018dust}, the production rate of the Trojan dust particles was calculated, and a numerical model was developed for the orbital motion of Trojan dust including the forces mentioned above. The evolution of these particles in the size range of [0.5, 32] $\mu\mathrm{m}$ (radius) was simulated, and the configuration of Trojan dust in form of arcs was derived \cite{liu2018dust} \textbf{(see Figure \ref{fig:SimuPolar})}. It was also found that the simulated number density is consistent with the upper limit obtained from the COBE data \cite{kuchner2000search}. Because the dynamics of the Trojan dust is affected by non-gravitational perturbation forces, including solar radiation pressure, solar wind drag, Poynting-Robertson drag and the solar wind Lorentz force, the distribution of the orbital parameters of the Trojan dust differs from that of the Trojan asteroids \cite{liu2018comparison}. Especially, in that study \cite{liu2018comparison}, the bimodal pattern of the semi-major axis distribution for Trojan dust was revealed. It was also shown that the distribution peak of the longitude of pericenter for the Trojan asteroids is about 60 degrees larger than Jupiter's longitude of pericenter, while for Trojan dust this difference in the longitude of pericenter is less than 60 degrees. The dynamical behavior of 2 $\mu\mathrm{m}$ (diameter) particles trapped in the Jupiter 1:1 resonance was analyzed in \cite{liou1995radiation, liou1995asteroidal}. A collisional numerical code was used to simulate the evolution of the Trojan dust particles with diameters in the range of [5, 500] $\mu\mathrm{m}$, including the effect of the Poynting-Robertson drag \cite{de2010studying}. Zimmer and Grogen integrated the orbital evolution of Trojan dust particles between 10 and 90 $\mu\mathrm{m}$ (diameter) for 500,000 years, and analyzed their fate \cite{zimmer2014orbital}. They considered the solar gravity, planetary gravitational perturbations, solar radiation pressure, Poynting-Robertson drag and solar wind drag. 

\section{Conclusions}
\noindent In the 2020s, there are two Jupiter missions and one Trojan mission scheduled for launch: ESA's JUpiter ICy moons Explorer mission, as well as NASA's Europa Clipper mission and Lucy mission to the Trojan asteroids. Especially onboard Europa Clipper, there is a dust instrument, the SUrface Dust Analyzer. An understanding of the Jovian dust environment will be important for the implementation of these missions, and in turn they will inform the preliminary modeling effects. In this review paper, we \textbf{have summarized} the results from previous observations and modeling work for Jovian dust outside the ring system including: particles in the region of the Galilean moons, Europa's dust plume, Jovian stream particles, grains in the outer region of the Jovian system, and dust in the region of the Jupiter Trojans. All relevant perturbation forces, the orbital evolution, and the spatial distributions of dust \textbf{were discussed}.

\subsection*{Acknowledgements}
\noindent This work was supported by the European Space Agency under the project ``Jovian Micrometeoroid Environment Model" (JMEM) (contract number: 4000107249/12/NL/AF) at the University of Oulu, and by the Academy of Finland under the project ``Earth and Near-Space System and Environmental Change".\\


\end{document}